\documentclass{llncs}
\usepackage{makeidx}  
\begin{document}
\frontmatter          
\pagestyle{headings}  
\mainmatter              
\title{{Mind Your Language}: Effects of Spoken Query Formulation on Retrieval Effectiveness}
%
%
\author{Apoorv Narang \and Srikanta Bedathur}
%
%
%
 \institute{IIIT-Delhi, Okhla Phase 3, New Delhi - 110020, India}

\maketitle              

\begin{abstract}
Voice search is becoming a popular mode for interacting with
search engines. As a result, research has gone into building better voice 
transcription engines, interfaces, and search engines that better
handle inherent verbosity of queries. However, when one considers its
use by non-native speakers of English, another aspect that becomes
important is the \emph{formulation of the query} by users. In this
paper, we present the results of a preliminary study that we conducted
with non-native English speakers who formulate queries for given
retrieval tasks. Our results show that the current search engines are
sensitive in their rankings to the query formulation, and thus
highlights the need for developing more robust ranking methods.
\keywords{Voice Search, Query Formulation, Evaluation}
\end{abstract}
\section{Introduction}

With the maturation of automatic speech recognition (ASR), voice
search systems have presented a new interface for information
retrieval. These voice search systems transcribe spoken queries and
use the text output for retrieval. However, as shown by Crestani
et. al \cite{cres}, spoken queries tend to be longer and more
natural. Also, even though ASR systems are improving rapidly, it has
been shown \cite{jiang} that transcription errors greatly influence
the performance of voice search systems.

Along with these challenges, search engines also need to adapt to the
variations in spoken query formulations. Compared to desktop search
queries, spoken queries can be more varied and loosely structured as
people begin to use natural language. While recent research has vastly
improved the transcription quality of ASR frontend of search engines,
as well as their handling of verbosity in ranking, it is not clear how
sensitive are the rankings to the linguistic structure of the spoken
query itself. This is particulary interesting for non-native English
speaking users of voice search.

In this paper, we present the preliminary results of our study which
involved a number of users who had training in English as foreign
language (EFL). Using a set of standard TREC topics, we first studied
if there is a difference in the way these users naturally formulate
their queries for the information need with a more well-formed
sentence as given in the TREC deescriptions themselves. Next, we
evaluated the effectiveness of results returned by Google - the most
popular Web search engine which provides an easy voice search
interface for us to experiment with. Our results show that although
search engines, as exemplified by Google, are very good in handling
various speech artefacts and verbose queries, their rankings are quite
sensitive to the query formulations. We observed a reduction of
20-30\% in the ranks where the most relevant results were shown as a
result of this.



\section{Setup}

Our major objective was to compare how the search engine performed
when users formulated their own spoken queries as compared to the
well-structured queries for the same topics.

For our evaluation, we used Google's voice search app on iOS with 20
TREC topics from the Web Track. Another advantage of using Google's
voice search system was that it is already believed to be tuned to
conversational queries with the new 'Hummingbird' algorithm. This
would give us a greater clarity of whether these state-of-the-art
systems are able to handle variations in query formulation compared to
well-structured queries.

Though some studies show that most mobile voice search queries are
local, similar to Jiang et. al \cite{jiang}, we didn't want to
restrict ourselves to just local queries because our experiment didn't
simulate mobile conditions and voice search systems are being used on
the desktop as well. Thus, we used 20 informational queries from TREC
Web Track in 2010, 2011 and 2012. Table 1 shows the list of these
topics. 

\begin{table}
\caption{TREC Web Track topics chosen for the experiment}
\begin{center}
\begin{tabular}{r@{\quad}rl}
\hline
\multicolumn{1}{l}{\rule{0pt}{12pt}
                   Year}&\multicolumn{2}{l}{Topic numbers}\\[2pt]
\hline\rule{0pt}{12pt}
2010  & 54, 55, 58, 69, 71, 74, 81 & \\
2011  &  110, 117, 125, 130, 131, 142  & \\
2012 & 157, 161, 166, 170, 175, 180, 181 & \\[2pt]
\hline
\end{tabular}
\end{center}
\end{table}

For the experiments, we used the latest version of Google's voice
search app for iOS set up to transcribe Indian English. We created new
Google accounts for each one of our participants with the 'Web
History' setting switched on to record their transcribed queries as
well as their clicks, based on which we calculated Mean Reciprocal
Rank for each query. 

Our experiment was conducted in two stages with these topics. We
called in 13 participants (9 males \& 4 females), all students in
higher education who have received formal education in English as
foreign language throughout their life. In the first stage, we gave
them each of the 20 TREC topics along with the information need, and
then asked them to formulate their own voice query. They then explored
the results while their 'clicks' were being recorded.  

After the first stage was over will all participants, we called them
in for the second stage on the experiment. In this stage, we gave them
a well-structured query in the form of 'Description' for each TREC
topic from the Web Track. The participants then spoke these queries
into the Google Voice Search app and again browsed the results.  

It was important to conduct this stage after the first one so that
participants aren't exposed to well-structured queries for the same
topics beforehand, thus influencing their queries. We also ensured
that participants do not type any queries and only speak them into the
app. Throughout the experiments, we allowed participants to correct
voice transcription errors, if any, while keeping a record of these
errors. We later used the knowledge of these transcription errors in
our evaluation to calculate 'best' and 'worst' MRR scores for each
spoken query.

\section{Evaluation Results}

In this section, we present the results of our experiments by focusing
mainly on the Reciprocal Rank (MRR) measure. The reciprocal rank is
simply the reciprocal of the position of the first result that was
marked relevant by the user in the rankings. In the perfect ranking of
results by a search engine, this should be $1$. 

In table~\ref{tab:alltrec}, we show the summary of performance for each of the 20
TREC queries we considered in this study. It shows RR values under
four different settings -- first two columns show the results for
queries that were naturally formulated by users, and the next two show
the results when TREC description queries were spoken by the same
users. For each of these two queries, we also show the worst
reciprocal rank obtained -- to account for the transcription errors.

These results show many interesting aspects: first of all, as recent
results have also shown, transcription errors do play an important
role in the quality of results. There is difference of about $0.12$ in
the MRR values with and without transcription, even with TREC
queries. At the same time, equally strong is the effect of query
formulations themselves. Specifically, the MRR value for TREC queries
is $0.9$ while for the naturally spoken queries for the same topics
have only $0.76$ which is much more than the reduction in quality due
to transcription errors alone.

We also highlight that these reductions are, though consistent, are
more pronounced in some topics and for some users. We illustrate this
point further by considering only those users whose natural queries
yielded low MRR values, and compare the MRR values for the same users
when they spoke TREC queries to the search system. These results are
shown in table~\ref{tab:lowMRR}. As these results show, there is a
significant reduction in the quality of rankings when users are
allowed for formulate their own queries, even when there are no errors
due to transcriptions alone.

\begin{table}
\caption{MRR for all TREC queries}
\begin{center}
\begin{tabular}{|l| c @{\quad} c | c @{\quad} c|}
\hline\rule{0pt}{12pt}

Topic \#         & \multicolumn{2}{| c |}{Natural Queries} &
\multicolumn{2}{| c |}{TREC queries}\\
&  RR & Worst RR & RR & Worst RR \\
\hline
54           & 0.72 &	0.65	& 0.79	& 0.79 \\
55            & 0.65 &	0.61 &	0.96 &	0.92 \\
58            & 0.82 &	0.75	& 1.00 &	0.83  \\
69            & 0.77 &	0.73 &	0.92 &	0.67  \\
71            &  0.54 &	0.52 &	0.85 &	0.73 \\
74            & 0.96 &	0.83 &	1.00 &	0.92  \\
81            & 0.74 &	0.55 &	1.00 &	0.82 \\
110            & 0.43 &	0.37	& 0.47 &	0.39 \\
117            & 0.69 &	0.46 &	0.69 &	0.43 \\
125            & 0.96 &	0.88 &	1.00 &	0.95  \\
130            & 0.88 &	0.72 &	1.00 &	0.88 \\
131            & 0.89 &	0.72 &	0.92 &	0.58  \\
142            & 0.67 &	0.67 &	0.91 &	0.68  \\
157            & 0.92 &	0.92 &	0.96 &	0.83  \\
161            & 0.64 &	0.53 &	0.86 &	0.62  \\
166            & 0.58 &	0.58 &	0.88 &	0.38 \\
170            & 0.81 &	0.67	& 0.78 &	0.61  \\
175            & 0.95 &	0.91 &	1.00 &	1.00  \\
180            & 0.87 &	0.87 &	0.92 &	0.50  \\
181            & 0.72 &	0.58 &	1.00 &	0.77  \\
\hline
{\bf MRR}      & 0.76 &	0.68 &	0.90 &	0.72  \\
\hline
\end{tabular}
\end{center}
\label{tab:alltrec}
\end{table}


\begin{table}
\caption{Avg. MRR for users with low MRR}
\begin{center}
\begin{tabular}{|l |  c | c |}
\hline\rule{0pt}{12pt}
No. of users         & Natural Queries & TREC queries  \\
\hline\rule{0pt}{12pt}
8            & 0.715 & 0.911  \\
\hline
\end{tabular}
\end{center}
\label{tab:lowMRR}
\end{table}

\begin{table}
\caption{Examples of Natural Queries with Low Result Quality}
\begin{center}
\begin{tabular}{|p{4.5cm}|p{6.5cm}|}
\hline
TREC Query         & Natural Query  \\
\hline
``Find information about the war in Afghanistan''   & ``get me some
information about the war in afghanistan'', ``tell me about the war
history of afghanistan'', ``history of afganistan wars''  \\
\hline
``I want to buy a road map of Brazil'' & ``i want to buy brazil's
map'',  ``i want to buy a map of brazil'', ``i want to buy a printed
map of brazil'', ``from where can i purchase map of brazil'',
``shopping results for map of brazil'', ``buy brazil map''  \\
\hline
``Find information about the office of President of the United States'' &
``give me some information about the current president of u s a'', ``who is
u s president'', ``tell me something about the president of the u s
a'' \\
\hline
\end{tabular}
\end{center}
\end{table}

\section{Conclusion}
In this paper we presented the preliminary results of our study in
understanding the current state of modern search engines in supporting
voice queries from a wide range of users. The results, though
preliminary and were conducted on a population of users who had much
higher levels of EFL training, show that there is a significant gap in
the performance of search engines for queries which are spontaneously
formed by users. This gap is as significant, and sometimes more than
the gap observed due to trasncription errors alone.

In future, we would like to expand our study to include users with
different levels of EFL training as well as wider range of queries. In
addition, we are also interested in developing improved search systems
which are robust for these artefacts.

%
%

%
%


\begin{thebibliography}{5}
%



\bibitem {cres}
Crestani, F., Du, H.:
Written Versus Spoken Queries: A Qualitative and Quantitative Comparative Analysis.
J. Am. Soc. Inf. Sci., 57: 881–890. (2006)

\bibitem {jiang}
Jiang, J., Jeng, W., He, D.:
How Do Users Respond to Voice Input Errors? Lexical and Phonetic Query Reformulation in Voice Search.
Proceedings of the 36th international ACM SIGIR conference on Research and development in information  retrieval (SIGIR’13), (2013)

\bibitem {google}
Schalkwyk, J., Beeferman, D., Beaufays, F., Byrne, B., Chelba, C., Cohen, M., Strope, B:
“Your Word is my Command”: Google Search by Voice: A Case Study.
Advances in Speech Recognition, pp. 61-90. Springer US (2010)

\end{thebibliography}
\end{document}